\documentclass[universe,article,accept,pdftex,moreauthors]{Definitions/mdpi} 

\firstpage{1} 
\pubvolume{1}
\issuenum{1}
\articlenumber{0}
\pubyear{2024}
\copyrightyear{2024}
\datereceived{ } 
\daterevised{ } 
\dateaccepted{ } 
\datepublished{ } 
\hreflink{https://doi.org/} 

\usepackage{graphicx}
\usepackage{xspace}

\newcommand{\Teff}{$T_{\rm eff}$}
\newcommand{\lgg}{log\,$g$}

\def\vsini{$v\,$sin\,$i$\xspace}
\def\mx{MX~TrA\xspace}

\def\arxivprefixesep{:}

\newcommand{\eprint}[2][]{%
	{\tt\if!#1!#2\else#1\arxivprefixesep\ignorespaces#2\fi}%
}	

\bibpunct{[}{]}{, }{n}{}{,}

\Title{Modelling the TESS light curve of Ap Si star \mx }

\TitleCitation{Modelling of light curve of star \mx}

\Author{Yury Pakhomov $^{1,\dagger,\ddagger}$\orcidA{}*, Ilya Potravnov $^{1}$\orcidC{}, Anna Romanovskaya $^{1}$\orcidB{}, and Tatiana Ryabchikova $^{1}$\orcidC{}}

\AuthorNames{Yu.~Pakhomov, I.~Potravnov,  A.~Romanovskaya, and T.~Ryabchikova}

\AuthorCitation{Lastname, F.; Lastname, F.; Lastname, F.}

\address{%
$^{1}$ \quad Institute of Astronomy of Russain Academy of Sciences; 
}

\corres{Correspondence: pakhomov@inasan.ru}

\firstnote{Current address: Pyatnitskaya str., 48, Moscow, 119017 Russia.}  

\abstract{
	The TESS light curve of the silicon Ap star \mx was modelled using the observational surface distribution of silicon, iron, helium, and chromium obtained previously with the Doppler Imaging technique. The theoretical light curve was calculated using a grid of synthetic fluxes from line-by-line stellar atmosphere models with individual chemical abundances. The observational TESS light curve was fitted by synthetic one with an accuracy better than 0.001~mag. The influence of Si and Fe abundance stratification on the amplitude of variability was estimated. Also the wavelength dependence of the photometric amplitude and phase of the maximum light was modelled showing the typical for Ap Si stars behaviour with increased amplitude and anti-phase variability in far ultraviolet caused by the flux redistribution.  
}



\begin{document}

\section{Introduction}

Chemically peculiar magnetic Ap/Bp stars are characterized by strongly inhomogeneous distribution of chemical elements in their atmospheres, both over the surface and in depth. Driven by selective atomic diffusion \citep{Michaud_1970} these inhomogeneities follow the surface geometry of magnetic field and form the horizontal abundance gradients - the chemical spots with increased or decreased abundance of certain elements up to few dex (in logarithmic scale) relative to the Sun. Due to the line blanketing the opacity differs significantly within spots and quiet photosphere that affects the emergent flux.  
The axial rotation of a spotted star leads to periodic brightness variability.
This is the key ingredient of the "oblique rotator"\, model \citep{1950MNRAS.110..395S}, which successfully explain variability of Ap/Bp stars.

Recent development in computation of model atmospheres with individual chemical composition \citep{Shulyak_2004} as well as observational Doppler imaging (DI) technique \citep[see e.g.][for review]{Piskunov_1993, Kochukhov_2016} has led to major advances in the interpretation of observations of Ap/Bp stars. In their study of model atmosphere computed with line-by-line opacity treatment \citet{2007A&A...469.1083K} showed that individual abundance patterns result in changing the atmospheric structure: temperature and pressure distribution, and modify the spectral energy distribution (SED). Elements such as silicon, iron and chromium which are often significantly overabundant in the line forming region of Ap/Bp atmospheres, were found to be the principal contributors to opacity. In particular, Si plays an exceptional role both in line and continuum opacities in the ultraviolet (UV) region \citep{Artru_1987}  that leads to the flux redistribution between far- ($\lambda \lesssim 1600$\,\AA\,) and near-UV as well as visible spectral region. In turn, this effect manifests in distinct photometric behaviour of silicon Ap/Bp stars depending on the bandpass. {The combination} of these theoretical and numerical advances with the capability of surface mapping with DI offers a direct opportunity for robust modelling and interpretation {the} spectral and photometric variability of Ap/Bp stars.

In a series of papers \citep{Krticka_2007,Krticka_2009,Shulyak_2010,Krticka_2012} the light curves of Ap/Bp stars were modelled using the surface distributions of elements obtained by DI method. As a result, a sufficiently good agreement between the synthetic light curves and the observed ones was obtained, proving surface elemental inhomogeneity as the reason for the light changes of investigated stars. Also, in agreement with theoretical predictions, the key contribution of spots with silicon, iron, and chromium overabundance to the rotational light modulation was confirmed.

Nevertheless, the problem is still relevant due to the growing number of precise high-cadence photometric observations from the spacecrafts Kepler/K2 \citep{Borucki_2010}, BRITE \citep{Weiss_2014}, TESS \citep{Ricker_2015} etc., and accurate representation of the light curves of Ap/Bp stars  may shed new light on some new effects. Thus, the large diversity of the surface distributions of elements in Ap/Bp stars will allow to disentangle the contribution of individual element to the total photometric variation and compare with theoretical expectations. Impact of effects such as deviation from the local thermodynamic equilibrium (LTE) and vertical abundance gradients (stratification) on the flux variations of Ap/Bp stars is not completely explored and its accounting will probably improve the accuracy of the fit to the observations.

The aim of the present work is the quantitative investigation of the effect of inhomogeneous surface abundance distribution of most peculiar elements in Ap star \mx on its photometric variability. \mx (HD~152564) is a bright southern Ap star which belongs to silicon subgroup of this class and demonstrates pronounced photometric variability \citep{Paunzen_1998}.  Based on high-resolution phase-resolved spectroscopy of \mx  \citet{2024MNRAS.52710376P} determined the fundamental and atmospheric parameters ({the effective temperature} \Teff\,=\,11950\,$\pm$\,200~K, {the surface gravity} \lgg\,=\,3.6\,$\pm$\,0.2, {the microturbulent velocity} $\xi_t$\,=\,0.0~km/s, {the macroturbulent velocity} $\zeta_{macro}$\,=\,0.0~km/s, {the projection of the rotational velocity} \vsini\,=\,69\,$\pm$\,2~km/s, {the mass} $M/M_{\odot}$\,=\,2.1, {the radius} $R/R_{\odot}$\,=\,3.8), abundances of 12 ions of 9 elements and also mapped the surface abundance distributions for Si, He, Fe, Mg, and O using DI. Later, \citet{2024INASR...9....1P} obtained Cr maps with the same DI technique. Analysis of high-precision photometric data from the TESS spacecraft revealed a period of $P=2.1639$~day, which, together with the values of the projectional rotational velocity \vsini and the radius of the star $R$, gives the value of the orbital inclination $i=51^{\circ}$. The phase light curve has an amplitude of about 0.03$^m$ and a quasi-sinusoidal shape, which reflects the modulation by the rotation of the spotted star. A pronounced rotationally modulated variability due to a highly inhomogeneous surface distribution of elements and expected presence of vertical stratification of silicon and iron makes \mx an exceptionally suitable object for investigation in the context of the problem above.

{The present work is based on the Doppler Imaging results obtained in \citet{2024MNRAS.52710376P, 2024INASR...9....1P} and focuses on the modelling of the light curve of \mx.}
The paper is organized as follows: Section~\ref{sec:obs} provides data on the photometric observations of \mx and surface abundance maps we used. Section~\ref{sec:model} describes the modelling technique, calculations of the surface intensity map and synthetic light curve, Section~\ref{sec:disc} describes the results and comparison with observations. In Section~\ref{sec:conc} we present our conclusions.

\section{Observational data}
\label{sec:obs}

{For construction of the light curve of \mx we used the photometric observations obtained by the TESS mission during its Cycle 3 in Sector 39 (from 05/27/2021 to 06/24/2021) and comprising about 20\,000 measurements with 120-s cadence in total. We used the photometry obtained during this cycle as being closest to the main period of the spectroscopic observations used for Doppler Imaging in \citet{2024MNRAS.52710376P}. Given the good representation of the \mx light curve from season to season, these observations are reliable for our main purpose of accessing the general shape and amplitude of the light curve.} The data were retrieved through the Mikulski Archive for Space Telescopes \footnote{https://mast.stsci.edu/portal/Mashup/Clients/Mast/Portal.html} (MAST) portal automatically processed with the Science Processing Operations Center (SPOC) software package \citep{Jenkins_2016}. {The processed SPOC light curve provides two types of fluxes: SAP (Simple Aperture Photometry) flux and PDCSAP (Pre-search Data Conditioning SAP) flux with removed long-term trends. In particular case of MX TrA photometry in Sector 39 PDCSAP fluxes after removing an offset appeared to be almost identical to the SAP fluxes but more noisy, hence we used latter for magnitudes conversion.} The SAP fluxes were converted to the TESS magnitude scale which is close to the Cousins $I_C$ filter using the formula $m_{TESS} = -2.5\,\mathrm{\mathbf{log}}(SAP\_FLUX) + 20.44$, taken from the project documentation \footnote{https://tess.mit.edu/public/tesstransients/pages/readme.html\#flux-calibration}.
The phased light curve of \mx was built using an ephemeris JD(max.light) = $2458647.7774 + 2^d.1639\,E$ from \citet{2024MNRAS.52710376P}. It should be noted that the light curve of \mx is slightly asymmetric, so the initial epoch in the ephemeris was determined from the centre of gravity of the maximum. This explains a small negative shift relative to zero of the points with maximal brightness in the phased light curve.

We also used surface distribution maps of 4 elements: He, Si, Fe from \citet{2024MNRAS.52710376P} and Cr from \citet{2024INASR...9....1P} for calculations of the synthetic light curves. These maps with a resolution of about $11^\circ$ on equator were obtained with DI technique based on a spectroscopic time series from the 10-m South African Large Telescope (SALT). Details of the spectroscopic observations and Doppler Imaging procedure are given in the papers cited above. For chromium, \citet{2024INASR...9....1P} provide two versions of the maps with slightly different abundance scale, depending on the set of lines they used. We explore both of them in attempt for the best representation of the observed ligth curve (see below).

\section{Modelling of light curve}
\label{sec:model}
Our approach to light curve modelling is based on the following fundamental steps: 
\begin{itemize}
 \item Construction of a surface intensity map. For this purpose, the specific intensities in the elements of surface grid are calculated with the individual abundances from the Doppler maps. 
 \item Disk integration of specific intensities for all rotational phases. Convolution of the integrated flux at each phase with the bandpass of the chosen filter allows to compare the calculated synthetic light curve with the observed one.
\end{itemize}
Details are given in the subsections below.

\subsection{Construction of intensity map}
\label{sec:map}

\begin{figure}
	\centering
	\includegraphics[clip,width=\columnwidth]{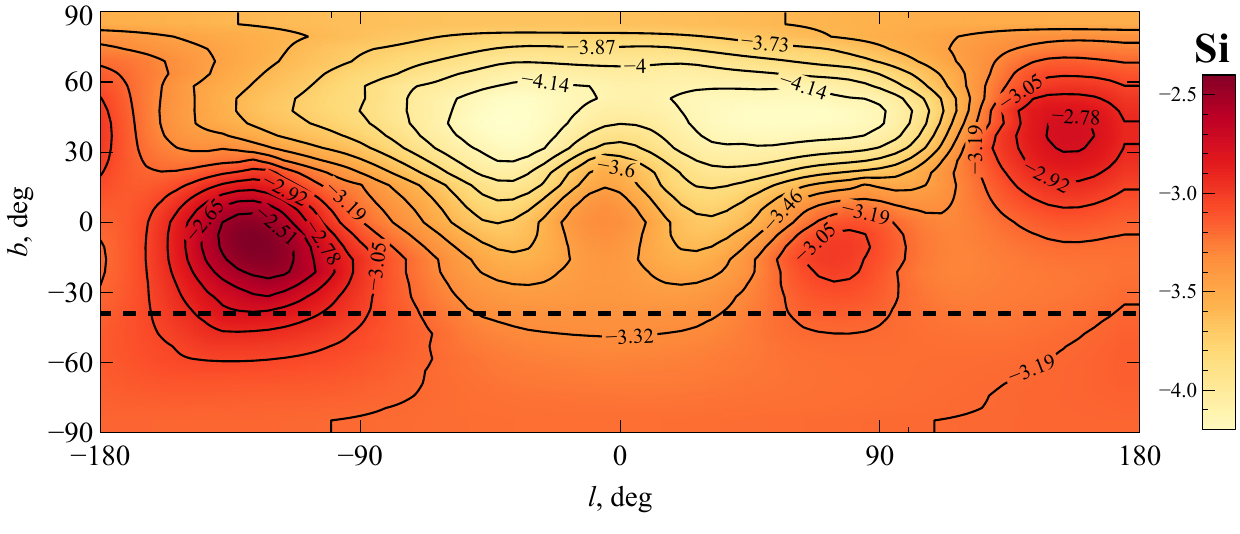}
	\includegraphics[clip,width=\columnwidth]{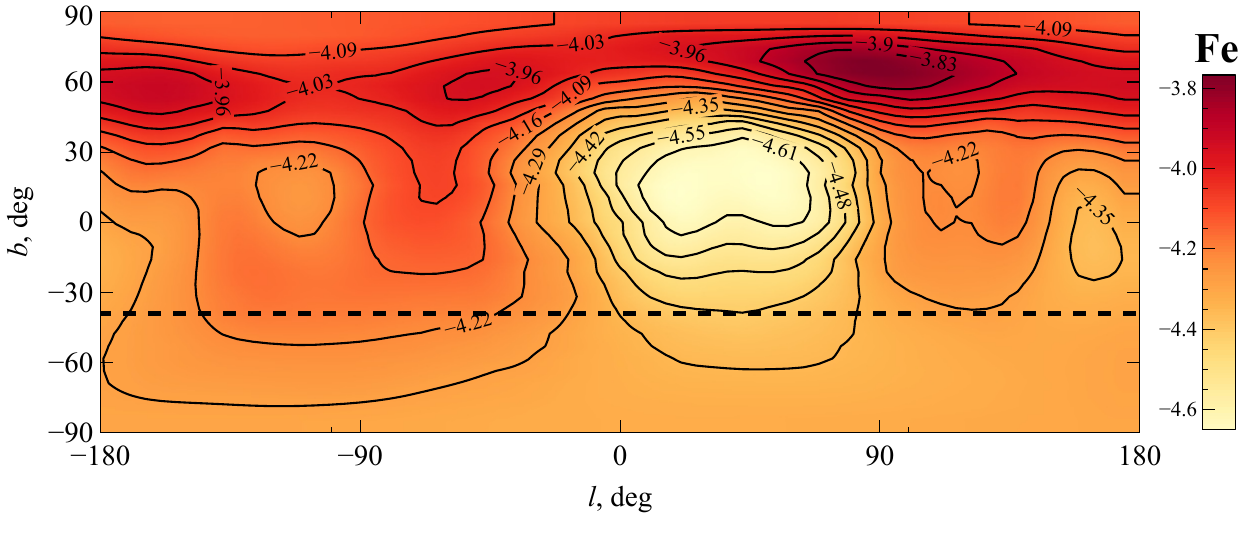}
	\includegraphics[clip,width=\columnwidth]{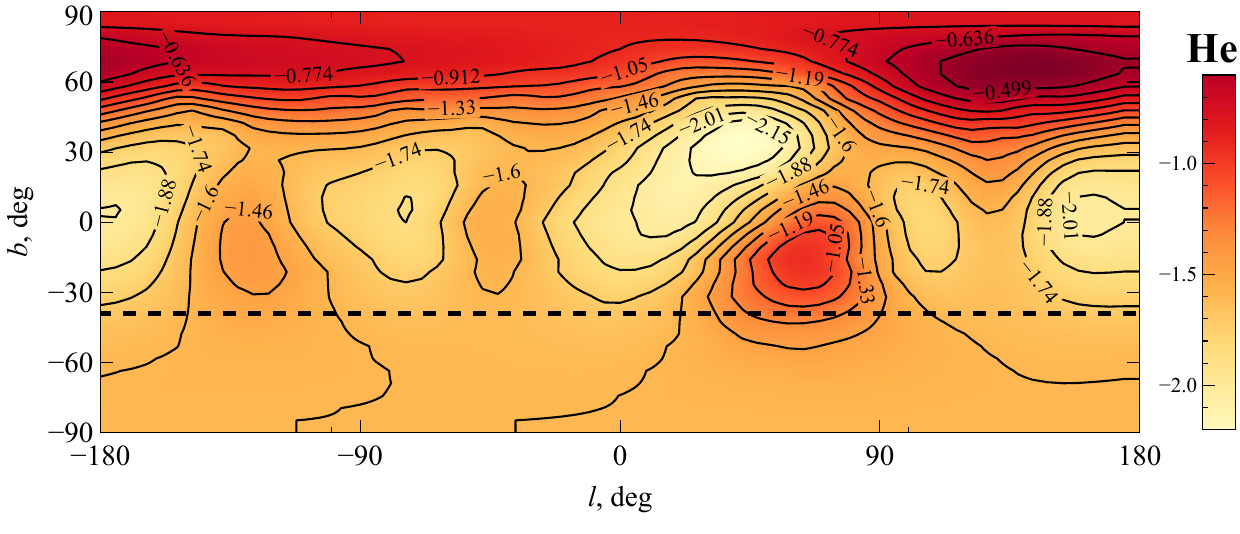}
	\includegraphics[clip,width=\columnwidth]{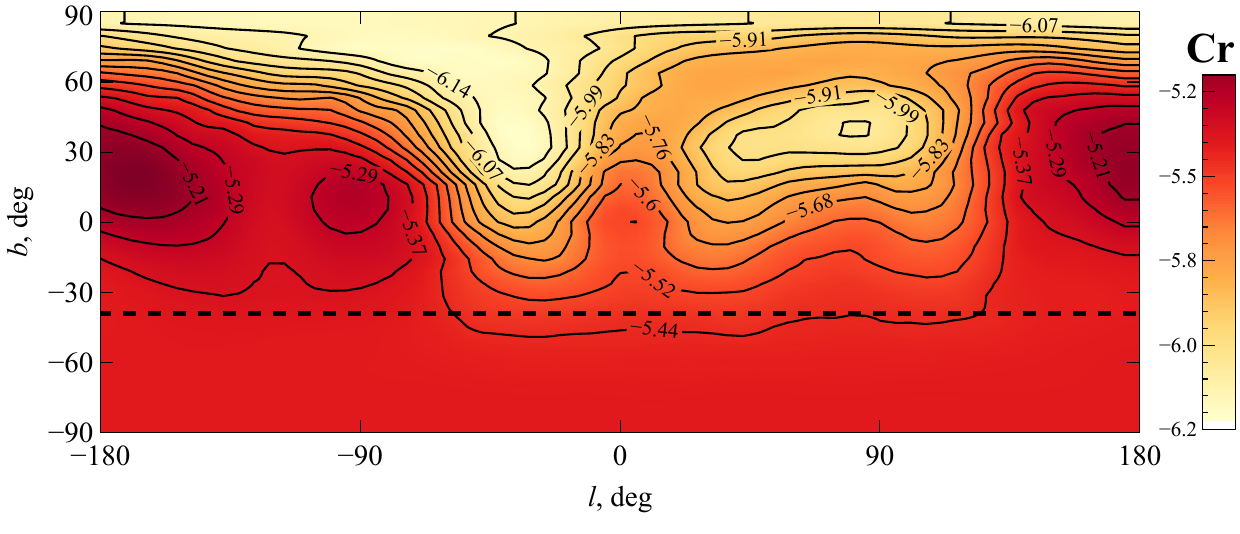}
	\caption{Maps of the distribution of silicon, iron, helium, and chromium (sequentially from top to bottom) on the surface of the star MX~TrA. The dashed line marks the lower boundary of the visible part of the surface due to the tilt of the rotation axis.}
	\label{fig:map}
\end{figure}

\begin{figure}[t]
	\includegraphics[clip,width=\columnwidth]{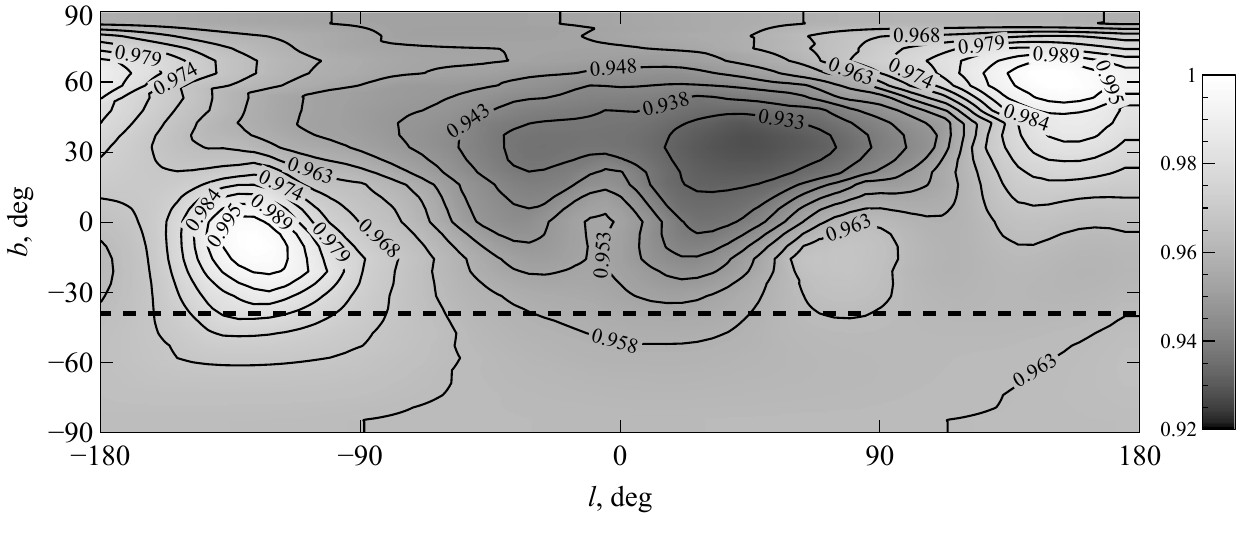}
	\caption{Map of the relative intensity distribution in the TESS image receiver bandpass. The maximum intensity value is taken as unit.}
	\label{fig:mapI}
\end{figure}

At the first step a grid of stellar atmosphere models was calculated with the LLmodels code \citep{Shulyak_2004} taking into account the individual chemical composition. The stellar parameters of \mx (\Teff\,=\,11950\,$\pm$\,200~K, \lgg\,=\,3.6\,$\pm$\,0.2) and the mean elemental abundances were adopted from \citet{2024MNRAS.52710376P}. The abundances of silicon, iron, helium, and chromium were varied within the limits inferred from Doppler maps (Fig.~\ref{fig:map}). The maps are 544$\times$272 pixels in size, which corresponds to a equidistant step of about 0.66$^{\circ}$ in latitude and longitude. The final grid consists of 256 atmosphere models calculated for all possible combinations of abundances in the ranges of $\log A_\mathrm{Si}$=[-4.50\, ... -2.30], $\log A_\mathrm{Fe}$=[-4.70\, ... -3.70], $\log A_\mathrm{He}$=[-2.11\, ... -1.61], and $\log A_\mathrm{Cr}$=[-6.50\, ... -4.10], where abundances $\log A_\mathrm{X} = \log(N_\mathrm{X}/N_\mathrm{tot})$ {are expressed through the ratio the number density $N_\mathrm{X}$ of element X to the total number density $N_\mathrm{tot}$}. The abundances of other elements remained unchanged. The synthetic SEDs were computed simultaneously with the atmosphere models. Using the response curve of the TESS imaging receiver \citep{2015JATIS...1a4003R} we calculated the flux and intensity of radiation from the 1~cm$^{2}$ of stellar surface. The calculated intensities in the TESS magnitude scale were combined into a grid with the gradients significantly smoothed out in the logarithmic scale. For each point in the surface map the specific intensity was calculated using the grid interpolation and taking into  account local abundances of Si, Fe, He, and Cr. Thus an intensity map $I_{l,b}$ was constructed in the bandpass of the TESS image receiver. The ratio of the minimum and maximum intensities was about 0.93. {This map scaled to the maximum value is shown in Fig.~\ref{fig:mapI}.} By its appearance the intensity map better resembles the silicon distribution map. This was expected, since silicon makes the most significant contribution to the absorption coefficient, especially in the UV range. Due to energy redistribution in the stellar spectrum, strong absorption in the UV leads to an increase in flux in the visible range. Therefore, dark spots with overabundance of silicon in Doppler map appear bright in the intensity map.

\subsection{Synthetic magnitudes and light curve}
\label{sec:synt}

The intensity map in rectangular coordinates was further transformed into a spherical one in orthographic projection taking into account inclination of the rotation axis $i$=51$^{\circ}$ \citep{2024MNRAS.52710376P}.

Apparent intensity $I'_{l,b}$ of an arbitrary surface element ("point"\, in the map) with coordinates $(l,b)$ and intensity $I_{l,b}$ towards the observer is
\begin{equation}\label{eq1}
 I'_{l,b} = I_{l,b} [1-c_1 (1-\mu)-c_2 (1-\mu)^2] \mu |cos(b)| \delta b \delta l
\end{equation}
where $\mu=cos(\vartheta)$, $\vartheta$ is the vertex angle between the direction from the center of the star toward the observer and a point on the stellar surface with coordinates $(l,b)$; $c_1=0.1816$ and $c_2=0.1651$ are limb darkening coefficients; $|cos(b)| \delta b \delta l$ is the area of surface element on the sphere. 

Limb darkening coefficients for quadratic law $I_\mu=I_0 [1-c_1 (1-\mu)-c_2 (1-\mu)^2]$
were calculated from the emergent radiation intensities $I_{\mu,\lambda}$ convolved with TESS bandpass $T_\lambda$ for seven values of $\mu$ using Levenberg-Marquardt method \citep{Levenberg_1944, Marquardt_1963} to approximate function
\begin{equation}\label{eq2}
\frac{I_\mu}{I_{0}} = \frac{\int I_{\mu,\lambda} T_\lambda d\lambda}{I_{0} \int T_\lambda d\lambda} = 1-c_1 (1-\mu)-c_2 (1-\mu)^2 
\end{equation}
The calculations were made for stellar atmosphere model with average abundances of Si, Fe, He, Cr. Impact of individual chemical composition on the limb darkening coefficients is small and does not exceed 0.1~--~0.2\%. 

The radiation flux in the bandpass of the TESS image receiver $F_{TESS}=\sum I'_{l,b}$ is obtained by integrating contribution of all points over the visible hemisphere of the star ($\vartheta < \pi/2$). The synthetic magnitude is 
\begin{equation}\label{eq3}
m ^{syn} = -2.5\, log\,F_{TESS}
\end{equation}

The magnitude zero point here is equal to zero, because it was already taken into account when calculating the specific intensities $I_{l,b}$ in magnitude scale to create the grid.

The magnitude $m^{syn}$ refers to the average radiation flux from an area of 1~cm$^{2}$ on the stellar surface. The apparent magnitude $m^{TESS}$ was calculated as
\begin{equation}\label{eq4}
 m^{TESS}=m^{syn}-5\,\mathrm{log}\,\frac{\theta}{2\times 2.06265\times 10^{11}},
\end{equation}

Here, $\theta$ is the angular diameter of the star in $\mu$as. We neglect interstellar extinction due to its smallness \citep{2024MNRAS.52710376P}.

The angular diameter $\theta=0.18297$~$\mu$as was calculated from the difference between the observed average magnitude over the rotation period $\overline{m^{obs}}$ and the synthetic one $\overline{m^{syn} }$:
\begin{equation}\label{eq5}
\theta=2 \times 2.06265\times 10^{11} \times 10^{-\frac{(\overline{m^{obs}}-\overline{m^{syn}})}{5}}  
\end{equation}

Combining this angular diameter with the distance to the star of 191$\pm$9~pc, obtained from inversion of Gaia DR3 parallax ($\pi=5.2209\,\pm\,0.2382$~$\mu$as) \citep{2023A&A...674A...1G}, the radius of \mx $3.8\,\pm\,0.2$~$R_\odot$ was obtained and found to be in very good agreement with spectroscopic determination by \citet{2024MNRAS.52710376P}.

\begin{figure}
	\includegraphics[clip,width=\columnwidth]{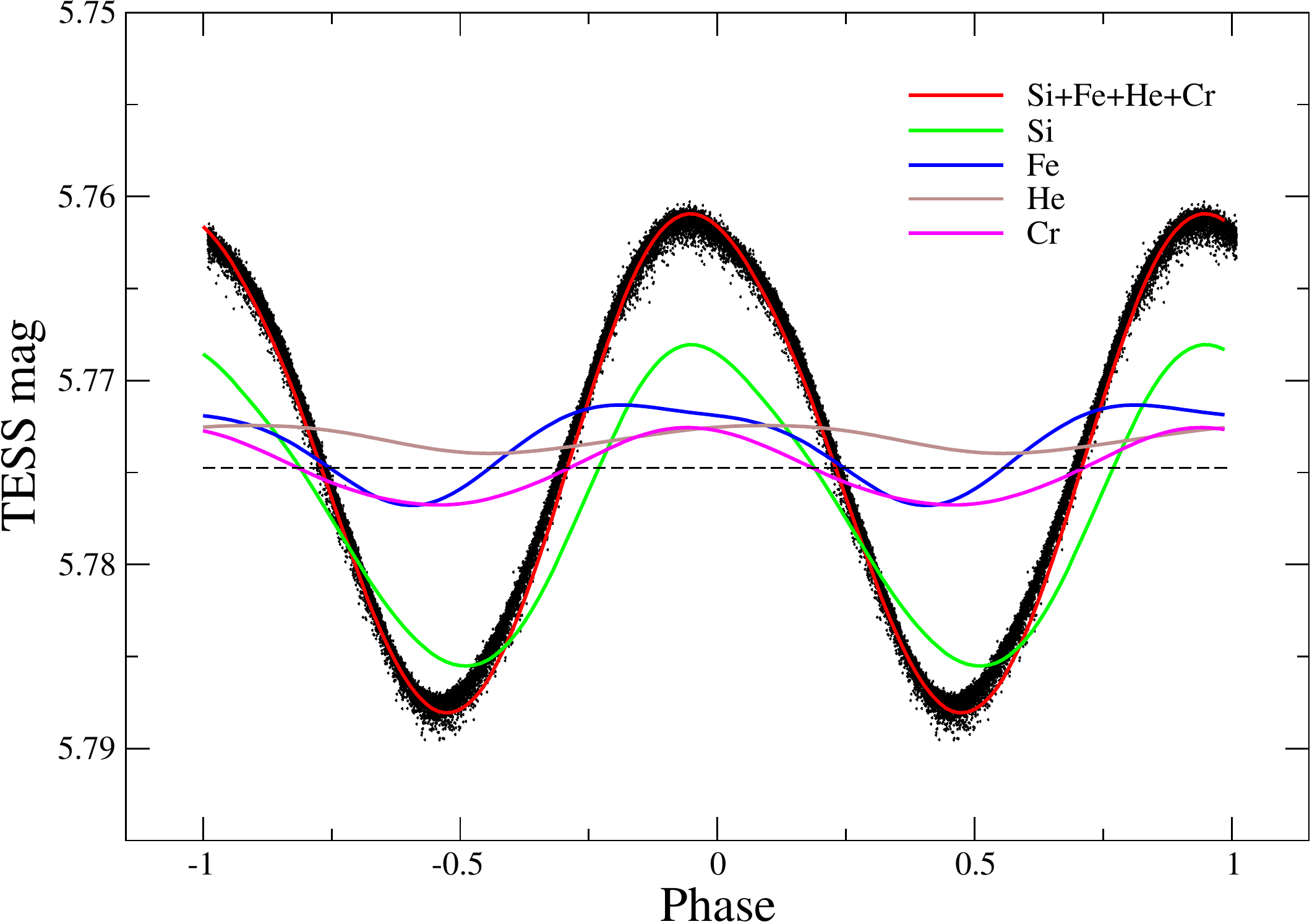}
	\caption{Comparison of the observed light curve of \mx with the theoretical ones. Each synthetic curve is calculated accounting for inhomogeneous surface abundance distribution of element(s). Dashed line indicates theoretical magnitude calculated for uniform surface abundance distribution with average values of elemental abundances.}
	\label{fig:curve}
\end{figure}

The synthetic light curve in $m^{TESS}$ units was computed with Eq.\ref{eq4} for the full set of rotational phases and is presented in Fig.~\ref{fig:curve} together with the TESS observations.

\section{Results and discussion}
\label{sec:disc}
\subsection{Synthetic light curve}
Fig.~\ref{fig:curve} represents theoretical light curves accounting for the individual contribution of each considered element and the total curve with cumulative impact of all elements in comparison with the observed TESS light curve. The observed light curve has a quasi-sinusoidal shape, with a more gradual descending {lag}. One can see from the figure that the total synthetic light curve perfectly matches the observations in terms of both amplitude and shape. Considering the individual contribution of elements, silicon has the largest one, about 64\%, to the amplitude of the light variation. The contributions at the phases of maximum and minimum are different due to inhomogeneous distribution of silicon spots over the stellar surface. That is why the light curve due to silicon surface abundance variations is asymmetrical relative to zero phase. This asymmetry is also manifested in the shape of the observed TESS light curve as the bar near phase $\varphi\approx0.1$. {A systematic error in Si abundance of the order of $\pm$0.2 dex, which is typical in abundance analysis, results in amplitude difference of order 0.005 mag. The amplitude reduces with silicon abundance decreases and increases with silicon in excess}. The next largest contributor to brightness variations is chromium with the value of relative amplitude about 22\%. Iron have amplitude about 20\%\ of total, but the light curve is shifted by phase which is consistent with the longitudinal position of the most contrast Fe spot from DI. Therefore, Fe provides somewhat lower contribution to total, about 17\%. Helium is responsible for the smallest changes in magnitudes. A feature in TESS light curve at phase of 0.1 is fitted well by Si, Fe, and He but with a somewhat reduced amplitude. Accounting for the contribution of chromium matches the amplitude, but the representation of light curve shape near maximum worsens due to ambiguity in chromium abundance scale. Exploring chromium maps with slightly different abundance scales from \citet{2024INASR...9....1P} reveals that the second map (their Fig.~2) with the lower abundance gradient provides the better fit of the light curve.

We also considered the possible contribution to the brightness variability of the light elements: magnesium and oxygen. These elements also possess a highly inhomogeneous surface distribution in \mx  \citep{2024MNRAS.52710376P}. The mean oxygen abundance is sub-solar (from NLTE analysis $\log A_\mathrm{O}\approx-4.0$), but the element is concentrated in three large equatorial spots with near solar abundance which occupy a significant fraction of the stellar surface. Magnesium is also depleted in \mx atmosphere with the mean abundance $\log A_\mathrm{Mg}\approx-5.0$ , but the region of its maximum (slightly sub-solar) abundance coincides with the circumpolar ring in the Fe distribution. We estimated the variability of TESS magnitudes due to inhomogeneous distributions of Mg and O computing intensity maps as described in Sect.~\ref{sec:map}. The differences between brightest and dimmest regions on the intensity maps are only 0.0001~mag and 0.0002~mag for Mg and O, respectively. Integration over stellar disc will significantly reduce these values. Therefore, the impact of these elements on brightness variations is negligible.  

In summary, inhomogeneous surface distribution of four elements: Si, Cr, Fe and He completely explains the observed photometric variations of \mx. This is in agreement with both theoretical expectations \citep{2007A&A...469.1083K} and modelling of light variations in other Ap/Bp stars \cite{Krticka_2007,Krticka_2009,Shulyak_2010}.  

\subsection{Estimation of impact of abundance stratification}

Vertical abundance gradients of elements (abundance stratification) affect the opacity distribution with depth in the atmosphere, resulting in differences in the emergent fluxes compared to a chemically homogeneous atmosphere. Therefore stratification should be considered as one of the effects potentially affecting the light curves of Ap/Bp stars. However, the straightforward accounting for stratification in light curve modelling is complicated, because the most suitable objects for stratification analysis are the stars with narrow spectral lines (low projected rotational velocities \vsini$\lesssim10-15$~km/s), but they appear inconvenient for DI and vice versa. \mx is no exception. Although its spectrum shows a large difference in abundances derived from spectral lines of different ionization stages of Si and Fe that is considered as evidence of stratification, the accurate reconstruction of the stratification profile is almost impossible due to the rapid axial rotation and severe line blending.

\begin{figure}
	\includegraphics[clip,width=\columnwidth]{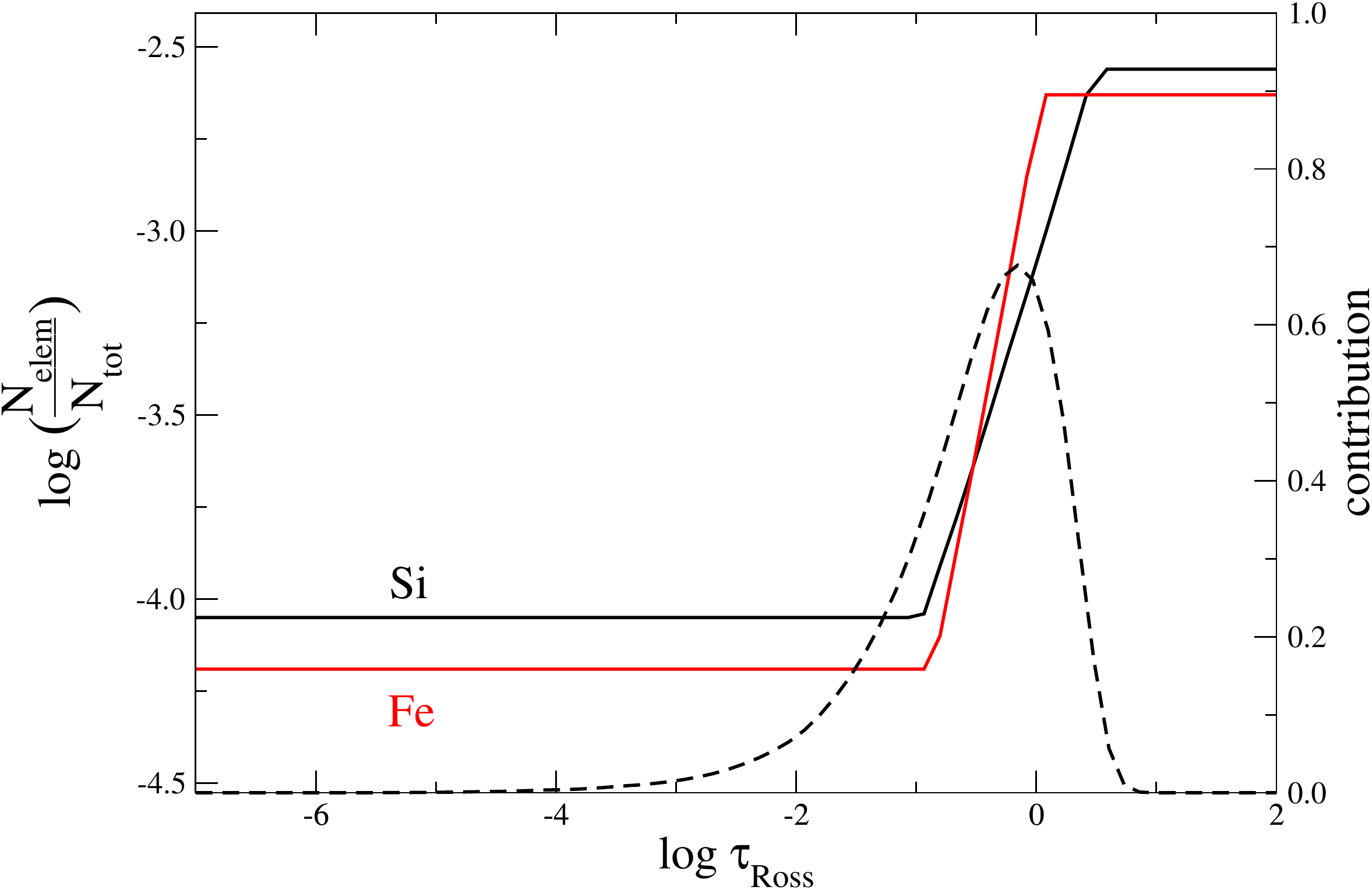}
	\caption{Stratification of silicon and iron in the atmosphere of the star BD+00$^\circ$1659  - twin of \mx. The {dashed} line indicates the wavelength-averaged function of the contribution of various layers to radiation in the visible range.}
	\label{fig:strat}
\end{figure}

Fortunately, in the \citet{Bailey_2013} list we found a star BD+00$^\circ$1659 which is a slowly rotating (\vsini=7~km/s) twin of \mx by its atmospheric parameters and chemical composition. A detailed analysis of this star, including stratification in its atmosphere will be presented in a forthcoming paper by \citet{Romanovskaya_2024}. In the present work we employ the stratification profiles for Si and Fe in BD+00$^\circ$1659 in the application to \mx. These profiles are presented in Fig.~\ref{fig:strat}, where the contribution function of the various atmospheric layers to the radiation in the TESS bandpass is also shown in scale of Rosseland optical depth. 

\begin{figure}
	\includegraphics[clip,width=\columnwidth]{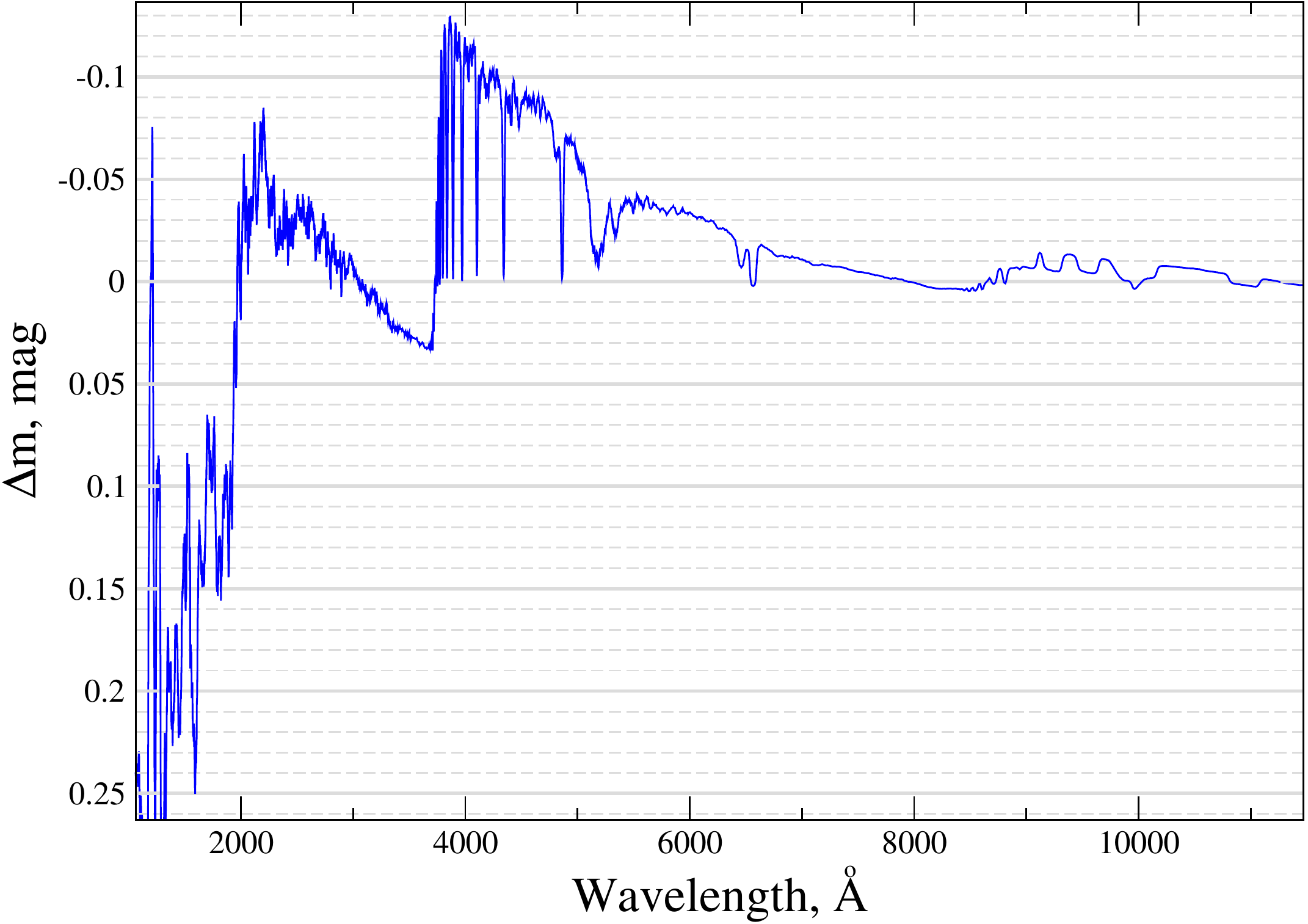}
	\caption{Wavelength dependence of flux difference for chemically homogeneous and {vertically} stratified atmosphere models.}
	\label{fig:dm}
\end{figure}

Two atmospheric models and corresponding synthetic SEDs were used to evaluate the effect: a chemically homogeneous model calculated for the mean abundances $\log A_\mathrm{He} = -1.60$~dex, $\log A_\mathrm{Si} = -3.46$~dex, $\log A_\mathrm{Fe} = -3.85$~dex and a stratified one calculated taking into account the abundance gradients of Si and Fe shown in Fig.~\ref{fig:strat}. The inhomogeneous horizontal distribution of elements was ignored at this step and these models were adopted for the entire atmosphere. This is justified by the fact that the current stratification analysis is not spatially resolved but is based on the radiation integrated over the visible hemisphere of the star. The difference in radiation fluxes between {vertically} stratified {($F_S$)} and chemically homogeneous {($F_0$)} models is $\Delta m = -2.5\log(F_S/F_0)$ and its wavelength dependence is shown in Fig.~\ref{fig:dm} {which referred to the disk-integrated flux with homogeneous horizontal distribution of chemical elements}. One can see that the maximum amplitude in the visible region reaches longward of the Balmer jump and the sign of the effect abruptly changes below $\lambda \lesssim 2000$~\AA. In the TESS bandpass the flux difference  reaches $-0.01^m$, i.e. in this case, stratification enhances the light amplitude. However, this is an upper limit. In reality stratified spots occupy  small fraction of the surface. We need to multiply the flux of stratified model on the filling factor $f\sim0.2$ corresponding to the fractional area of Si spots. Consequently, the flux ratio will be reduced by order of magnitude. Also the effect of stratification on emergent flux is very sensitive to the depth of the stratification step in the atmosphere as follows from the comparison with the contribution function in Fig.~\ref{fig:strat}. Shifting into upper atmosphere the step appears in a region (e.g. $\log \tau_{Ross}\approx-2$) where the contribution of layers to the continuum flux is small, thus reducing the difference in flux by up to two orders of magnitude. Depending on the position of stratification step the amplitude could both increase or decrease. The sophisticated 3D analysis with simultaneous accounting for both vertical and horizontal abundance gradients requires knowledge of the stratification profile in each surface element that is unavailable for rapidly rotating Ap stars like \mx. Generally, we estimate the contribution of vertical stratification to the light variations of \mx in the visual region as negligible, that is consistent with a good representation of the observations with horizontal abundance inhomogeneities only.

\subsection{UV variations}

\begin{figure}
	\includegraphics[clip,width=\columnwidth]{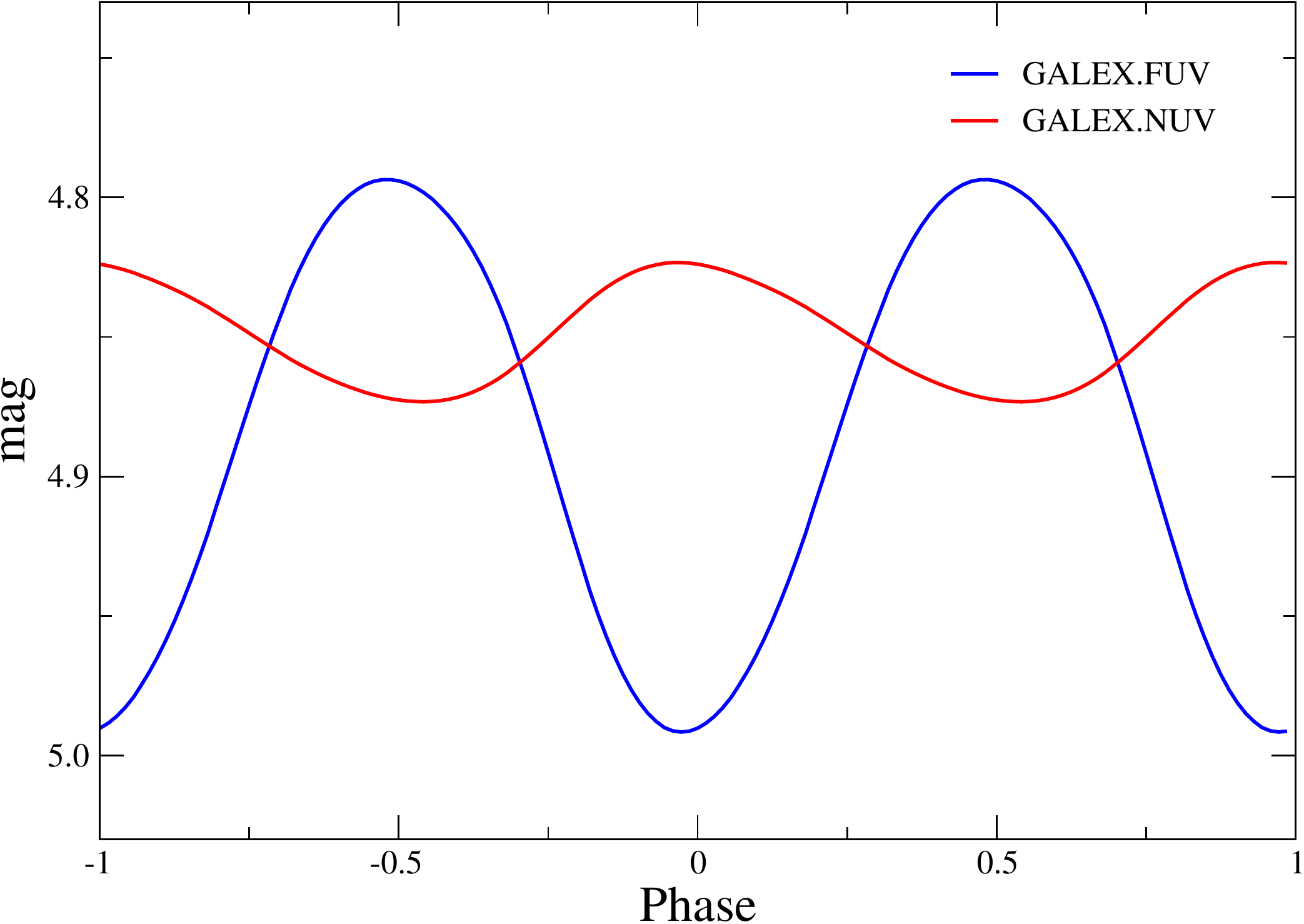}
	\caption{Predicted ultraviolet light curves {of \mx} for FUV and NUV bands of GALEX.}
	\label{fig:UV}
\end{figure}

One of the principal effects due to silicon overabundance in the atmosphere is the redistribution of flux between the far-UV and visible regions. Indeed, observations of some Ap Si stars clearly demonstrate the effect of phase shift or complete reversal of the light curve depending on the wavelength range \citep{Sokolov_2006,Shulyak_2010,Sokolov_2012}. Although photometric observations of \mx in the far-UV are yet not available, it is instructive to calculate synthetic light curve in this region (Fig.~\ref{fig:UV}). We used bandpasses of two GALEX filters for far-UV (FUV, centred at $\lambda\sim1530$~\AA\,) and near UV (NUV, at $\lambda\sim2350$~\AA\,)\footnote{https://asd.gsfc.nasa.gov/archive/galex/tools/Resolution\_Response/index.html}. Comparison of the two curves in Fig.~\ref{fig:UV} reveals the antiphase brightness changes in NUV and FUV filters while the NUV light curve is in phase with the visual TESS one. The amplitudes are also significantly different, with largest light variations in FUV. The physical basis for this difference is that bandpass of FUV filter centred shortward of 1527~\AA\ - the photoionization threshold of Si~I and contains numerous resonance lines and autoionization features of Si~II. Shortward ($\lambda \lesssim1600$~\AA\,) the energy is blocked due to absorption and redistributed to the longer wavelengths which leads to the flux increasing in near-UV and visual regions. This feature can be used for the photometric identification of Ap Si stars \citep{Jamar_1978,Romanovskaya_2021}.

\begin{figure}
	\includegraphics[clip,width=\columnwidth]{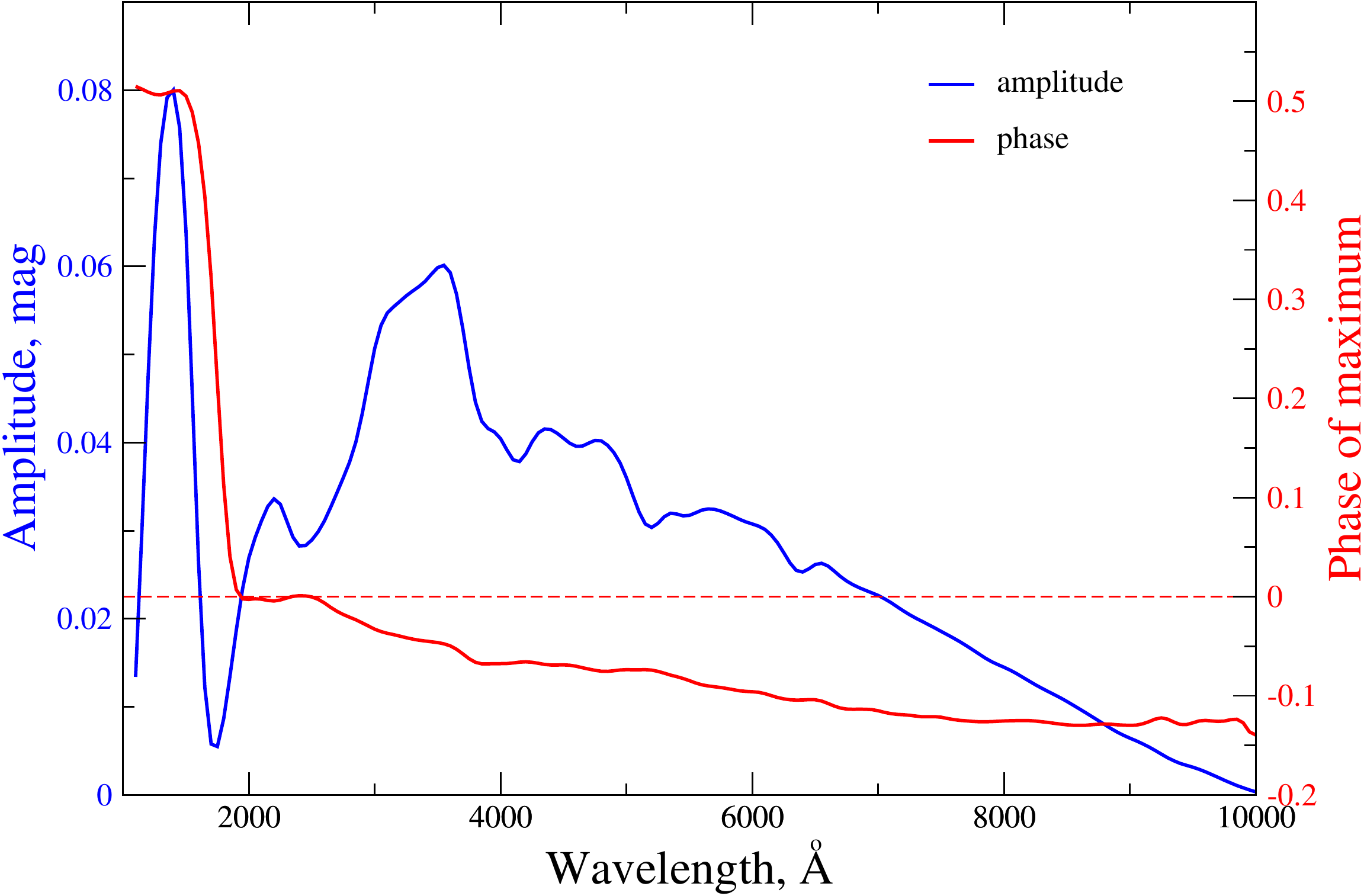}
	\caption{Wavelength dependence of amplitude and phase of the maximum of \mx photometric variability caused by inhomogeneous surface distribution of He, Si, Fe, Cr.}
	\label{fig:AmpPhase}
\end{figure}

Generalizing the approach, we calculated the light curves of \mx in the 1100-10000~\AA\, range with a 100~\AA\ filter and plot in Fig.~\ref{fig:AmpPhase} the wavelength dependence of the photometric amplitude and phase of the maximum. The figure clearly illustrates the amplitude increase toward the shorter wavelengths and the existence of a dip near 2000~\AA\, - the "null region"\, where the flux is almost constant over the rotational cycle. The existence of such a "null region (-s)"\, pointing to the mechanism of flux redistribution was previously detected in spectrophotometric observations of Ap stars \citep{Molnar_1973,Jamar_1978,Sokolov_2006,Sokolov_2012}. The phases of the maximum also differ on either side of "null region"\,. While at longward a gradual phase shift to negative values is expected, at short wavelengths where the flux is effectively blocked by silicon absorption there is a sharp increase up to a phase difference of $\Delta\varphi=0.5$ (anti-phase variability) relative the visual region.

\section{Conclusions}
\label{sec:conc}

In the present paper we report the results of modelling the high-precision TESS light curve of Ap Si star \mx based on the model of the oblique rotator and maps of surface elemental distribution previously obtained with DI technique. We were able to successfully reproduce the observed shape of the light curve and its amplitude with an accuracy better than 0.001~mag accounting for the inhomogeneous surface distribution of four elements: Si, Fe, Cr, and He. This list is enough for a good fit of the observations. Diversity of the surface distributions of elements leads to a phase shift and different contribution of an individual element to the light minimum and maximum. The total synthetic light curve perfectly reproduces the shape of the observed one. The contribution of light elements: O and Mg to the light variations appears to be negligible.

We also estimated the effect of the vertical stratification of Si and Fe in the \mx atmosphere on the emergent flux. We show that, in principle, stratification can contribute to light variations increasing emergent flux near Balmer jump and reducing it in the far-UV. However, in the TESS bandpass, the total effect does not exceed $\sim$0.01~mag and will be reduced by an order of magnitude taking into account horizontal chemical inhomogeneities. Hence, it does not contribute significantly to the TESS light curve amplitude.
Empirically, we conclude that taking into account only the inhomogeneous horizontal abundance distribution of Si, Fe, Cr, and He is enough for good representation the observed light curve of \mx in TESS bandpass.

The wavelength dependence of the amplitude of \mx light variations and phase of the maximum was calculated from synthetic light curves. It shows the well known for other Ap Si stars effect of increasing amplitude and antiphase variability between far-UV and visible regions. This result clearly demonstrates the possibility for identification a new Ap Si stars e.g. using photometric observations in the far-UV with the upcoming Spektr-UF (WSO-UV) space mission \citep{Shustov_2021} and phase-correlated optical observations.

\vspace{6pt} 


\authorcontributions{
	Conceptualization, I.P. and T.R.; methodology, Yu.P.; software, Yu.P.; validation, Yu.P., T.R. and I.P.; formal analysis, Yu.P., I.P., and T.R.; investigation, Yu.P.; resources, A.R.; data curation, A.R. and Yu.P.; writing---original draft preparation, Yu.P. and I.P.; writing---review and editing, Yu.P., I.P. and T.R.; visualization, Yu.P.; supervision, I.P.; project administration, Yu.P.; funding acquisition, I.P. All authors have read and agreed to the published version of the manuscript.}

\funding{This research was funed by the grant of Russian Science Foundation \textnumero24-22-00237, https://rscf.ru/en/project/24-22-00237/.}

\informedconsent{Informed consent was obtained from all subjects involved in the study.}

\dataavailability{Dataset available on request from the authors.} 

\acknowledgments{We obtained the observed data of the TESS space mission and processed using the SPOC (Science Processing Operations Center) automatic software package and obtained through the portal MAST (Mikulski Archive for Space Telescopes).
We thank Denis Shulyak for your program LLmodels and useful tips.}

\conflictsofinterest{The authors declare no conflicts of interest.} 



\abbreviations{Abbreviations}{
The following abbreviations are used in this manuscript:\\

\noindent 
\begin{tabular}{@{}ll}
GALEX & GALaxy evolution EXplorer - NASA orbiting space telescope \\
DI & Doppler imaging \\
LTE & Local thermodynamic equilibrium \\
NLTE & Non local thermodynamic equilibrium \\
SED & Spectral energy distribution \\
TESS & Transiting Exoplanet Survey Satellite \\
UV & Ultraviolet \\
\end{tabular}
}

\reftitle{References}


\bibliographystyle{Definitions/mdpi}
\bibliography{paper}



\end{document}